\documentclass[12pt,a4paper,superscriptaddress,showpacs,pre,preprintvip]{revtex4}
\usepackage{graphicx}

\begin{document}

\newcommand{\EQ}{Eq.~}
\newcommand{\EQS}{Eqs.~}
\newcommand{\FIG}{Fig.~}
\newcommand{\FIGS}{Figs.~}
\newcommand{\SEC}{Sec.~}
\newcommand{\SECS}{Secs.~}

\title{Analysis of scale-free networks based on a threshold graph
with intrinsic vertex weights}
\author{Naoki Masuda}
\affiliation{Laboratory for Mathematical Neuroscience,
RIKEN Brain Science Institute, 2-1, Hirosawa, Wako, Saitama, 351-0198
Japan}
\affiliation{Aihara Complexity Modelling Project, ERATO, JST, 3-23-5, Uehara,
Shibuya, Tokyo, 151-0064 Japan}
\author{Hiroyoshi Miwa}
\affiliation{Department of Informatics, School of Science and
Technology, Kwansei Gakuin University,
2-1, Gakuen, Sanda, Hyogo, 669-1337 Japan}
\author{Norio Konno}
\affiliation{Faculty of Engineering,
Yokohama National University,
79-5, Tokiwadai, Hodogaya, Yokohama, 240-8501 Japan}
\date{Received 29 March 2004}

\begin{abstract}
Many real networks are complex and have power-law vertex degree
distribution, short diameter, and high clustering. We analyze the
network model based on thresholding of the summed vertex weights,
which belongs to the class of networks proposed by Caldarelli {\it et
al.} [Phys. Rev. Lett. {\bf 89}, 258702 (2002)].
Power-law degree distributions, particularly with the
dynamically stable scaling exponent 2, realistic clustering, and short
path lengths are produced for many types of weight
distributions. Thresholding mechanisms can underlie a family of real
complex networks that is characterized by cooperativeness and the
baseline scaling exponent 2. It contrasts with the class of growth
models with preferential attachment, which is marked by
competitiveness and baseline scaling exponent 3.
\end{abstract}

\maketitle

\section{Introduction}\label{sec:introduction}

Complex networks have drawn increasing interests in various disciplines.
Recent studies have revealed that networks in the real world are far
from fairly regular or totally random.  Particularly, real
networks have small average shortest path length and
high clustering at the same time, whereas conventional graphs such as
lattices, trees, or the original random graphs are not equipped with
these properties at the same time \cite{SW,reviews}.

The average path length is denoted by $L$, and it more or less
characterizes the diameter of the graph. The average of the shortest
path length over all the pairs of vertices defines $L$.  The
clustering property can be locally evaluated by the vertex-wise
clustering coefficient, which is the number of connected triangles
containing the vertex in question, normalized by the maximal number of
possible triangles.  If the vertex degree, or the number of edges
emanating from a vertex, is $k$, the normalization constant becomes
$k(k-1)/2$. The clustering coefficient $C$ of the whole graph is the
local clustering coefficient averaged over all the vertices.  Watts
and Strogatz proposed the small-world networks that simultaneously
realize large $C$ and small $L$ \cite{SW}. However, the small-world
networks are short of the scaling property of vertex degree
distribution denoted by $p(k)$. Indeed, not all but many real networks
satisfy $p(k)\propto k^{-\gamma}$ typically with $2< \gamma < 3$
\cite{reviews}.  Then, Barab\'{a}si and Albert (BA) developed the network
model, which dynamically generates scale-free networks with
$\gamma=3$ \cite{reviews,Barabasi99}.  The fundamental devices in the
BA model are the network growth and the preferential attachment;
vertices are added one after another to the network, and edges are
more prone to be connected to vertices with larger $k$.  Various
scale-free networks including extensions of the BA model, such as
networks with dynamic edge rewiring \cite{Albert00prl,Vazquez02},
those with nonlinear preferential attachment \cite{Krapivsky}, those
with weights on edges \cite{Yook}, the fitness model \cite{Bianconi},
and the hierarchically and deterministically growing models
\cite{Barabasi01,Jung,Dorogovtsev02,Ravasz02,Ravasz03} have been
proposed.  These models largely yield more flexible values of
$\gamma$, which is restricted to 3 in the original BA
construction. Furthermore, modifications to reinforce the clustering
property, which the BA model lacks, have also been done.  A simple
solution is to embed a triangle-generating protocol into the BA model
\cite{Holme,Szabo03}.  Introduction of a node deactivation procedure
also enhances clustering \cite{Klemm0202,Vazquez03}.  Yet another
solution is appropriately designed versions of the hierarchical models
mentioned above \cite{Dorogovtsev02,Ravasz02,Ravasz03}.  In short,
large $C$, small $L$, and scale-free $p(k)$ can be simultaneously
realized by the modified BA models or by the hierarchical construction.
Both models rely on the combined effects of network growth and the
preferential attachment, although preferential attachment is not
explicitly implemented in the hierarchical networks.

Nevertheless, every network is not apparently growing. Networks can
experience structural changes that are relatively much faster than
network growth or aging processes.  In economical networks of
companies, friendship networks, peer-to-peer (P2P) networks, and
networks of computer programs, for example, it
seems natural that agents change their connectivity without
significant entries or leaves of members.  Therefore, there has been a
need for developing a nongrowing algorithm to generate realistic
networks.  In this regard, Caldarelli {\it et al}. proposed a class of
networks whose connections are determined by interactions of vertices
that are endowed with intrinsic weights
\cite{Caldarelli,Soderberg,Boguna}.  The vertex weight is considered
as a type of fitness
\cite{Bianconi,Caldarelli,Chung,Barrat,Newman_Goh}, which describes
the propensity of vertices to gain edges. It can be interpreted as
money, social skills, or personal influence in social networks,
activeness, the value of information attached to a vertex,
concentration or mass of some ingredients in chemical or biological
networks, or the vertex degree itself.  Surprisingly, scale-free
topology spontaneously emerges even with weight distributions without
power laws \cite{Caldarelli,Boguna}.

In this paper, we analyze a subclass of their model that is based on a
deterministic thresholding mechanism.  The connectivity between a pair
of vertices is determined by whether the sum of the weights of the pair
exceeds a given threshold.  Actually, this class of networks is
equivalent to the threshold graph in the graph theoretical context
\cite{Golumbic}, and we also discuss its consequence.

Despite the stochasticity and certain continuity in the real world,
thresholding that is more or less ``hard'' is often observed.
Although the correspondence to our
framework is not perfect, a common form of thresholding is that an
agent on a vertex determines its action or state based on the number
of neighbors taking a specific state. For example, propagation of
riots, fashions, and innovations are considered to be equipped with
thresholding mechanisms \cite{Granovetter78}. These phenomena have
been simulated by dynamic models, such as the threshold model for
social decision \cite{Granovetter78}, the minority games
\cite{minority}, the threshold voter models, and the threshold contact
processes \cite{Liggettbook99}.

In this paper, we show that a baseline power law
$p(k)\propto k^{-\gamma}$ with $\gamma=2$ rather than one with $\gamma=3$
\cite{Barabasi99} dominates this class of models and explore its cause
and consequence.  In \SEC\ref{sec:model}, we follow
Refs. \cite{Caldarelli,Boguna} to explain the network model, and calculate
fundamental quantities such as $p(k)$, $C$, and the measure for degree
correlation.  The results in \SEC\ref{sec:model} are applied to
various weight distribution functions in \SEC\ref{sec:examples},
extending the results for the exponential distributions in
Refs. 
\cite{Caldarelli,Boguna}.  Consequently, we find that the power law is
observed for a wide class of weight distributions.  In
\SEC\ref{sec:universality}, we argue that the power law with
$\gamma= 2$ is rather ubiquitous in the sense that it is a unique
stable degree distribution when a network evolves without growth.

\section{Model}\label{sec:model}

Let us start with
a set of $n$ vertices $V = \{v_1, v_2, \ldots, v_n\}$.
As introduced in Refs. \cite{Caldarelli,Boguna},
we assign to each $v_i$ ($1\le i\le n$) a weight $w_i\in {\bf R}$ that is
taken randomly and independently distributed as specified by a given
probability distribution function $f(w)$ on ${\bf R}$.
The weight
quantifies the potential for the vertex to be linked to other
vertices \cite{Bianconi,Caldarelli}.
We assume that the weight permits
additive operation.  Actually, the multipliable weights
\cite{Chung,Caldarelli,Boguna,Newman_Goh} 
can be easily reduced to the additive ones by taking the logarithm of $w$.
Let
\begin{equation}
F(w) = \int^w_{-\infty} f(w^{\prime}) dw^{\prime}
\end{equation}
be the cumulative distribution function, satisfying $\lim_{w\to
-\infty} F(w) = 0$ and $\lim_{w\to\infty} F(w) = 1$.  The set of edges
$E$ is defined by the thresholding rule with threshold
$\theta$: $E = \{ (v_i, v_j) ; w_i+w_j\ge \theta, i\neq j\}$.  We
focus on this specific case of more general framework
\cite{Caldarelli,Boguna}. This renders more mathematical analysis and
allows us to explore the consequence of vertex interactions based on
intrinsic weights.  The degree distribution $p(k)$, where $0\le k< n$
is the vertex degree, is readily calculated with the use of continuum
approximation corresponding to the thermodynamic limit
($n\to\infty$). However, we confine ourselves to a finite $n$, and the
limit $n\to\infty$ should be understood as approximation.  Putting the
upper limit of $k$ equal to $n$ instead of $n-1$, we obtain
\begin{equation}
k = n \int^{\infty}_{\theta - w} f(w^{\prime}) dw^{\prime}
= n \left[ 1 - F\left(\theta - w\right) \right],
\quad (0\le k\le n)
\label{eq:k(w)}
\end{equation}
and
\begin{equation}
p(k) = f(w)\frac{dw}{dk}
= \frac{f\left(\theta-F^{-1}\left( 1 - \frac{k}{n}\right)\right)}
{n f\left( F^{-1}\left(1 - \frac{k}{n}\right)\right)}.
\label{eq:p(k)}
\end{equation}

\begin{figure}
\begin{center}
\includegraphics[height=2.25in,width=3.25in]{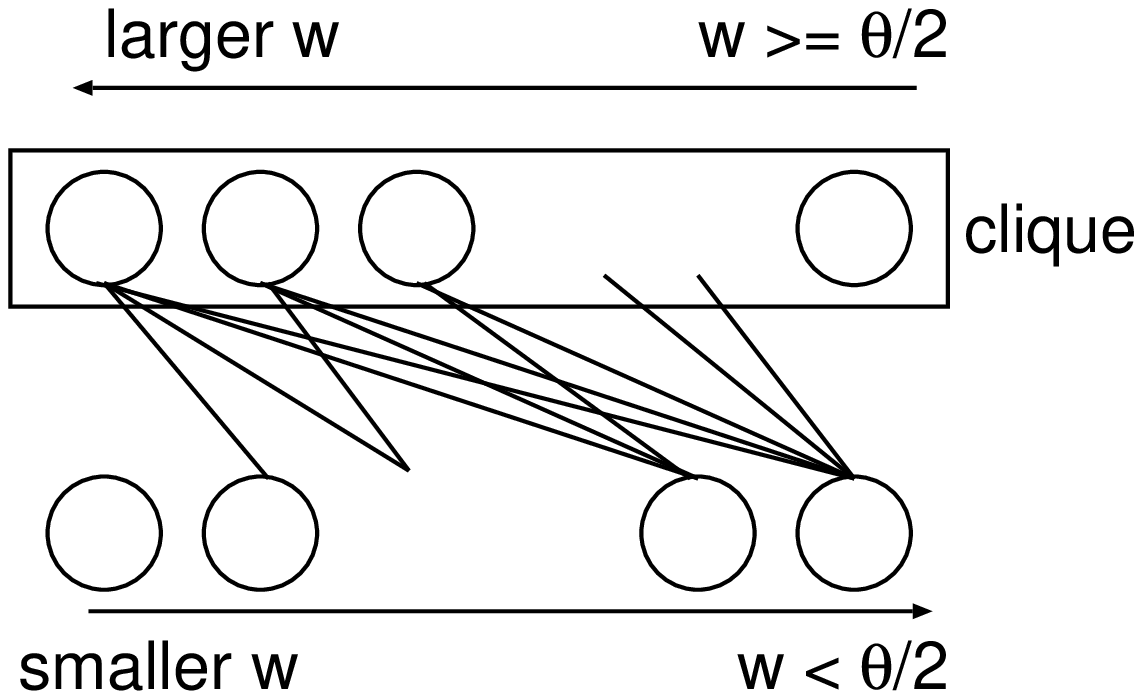}
\caption{}
\label{fig:separation}
\end{center}
\end{figure}

Because of the one-to-one correspondence between $k$ and $w$
represented by \EQ(\ref{eq:k(w)}), the vertex-wise cluster coefficient
depends only on $k$, which simplifies the analysis.
We denote it by $C(k)$, and the scaling law
$C(k)\propto k^{-1}$ is often observed in real and modeled networks
\cite{Dorogovtsev02,Ravasz02,Ravasz03,Klemm0202,Vazquez03,Szabo03,Newman03}.  The clustering
coefficient of the entire graph is given by $C = \int^{\infty}_0 C(k)
p(k) dk$.  To calculate $C(k)$ \cite{Boguna}, let us consider a vertex $v$ with
degree $k= n \left[ 1 - F(\theta - w) \right]$.  The density of the
number of neighbors with degree $k^{\prime} = n \left[ 1 - F(\theta -
w^{\prime}) \right]$ becomes $f(w^{\prime})$ if $w^{\prime}\ge \theta
- w$ and 0 otherwise. With such a neighbor denoted by $v^{\prime}$,
the number of connected triangles comprising $v$, $v^{\prime}$ and
another neighbor of $v$ is obtained as follows.  When $k^{\prime}\ge
k$, a new neighbor of $v$ is also a neighbor of $v^{\prime}$ because
$w^{\prime} \ge w$.  The number of triangles in this case is 
\begin{equation}
(n-2)
\left[ 1 - F(\theta - w) \right] \cong n \left[ 1 - F(\theta -
w) \right] = k.
\label{eq:k_w_1}
\end{equation}
When $k^{\prime}< k$, we have
\begin{equation}
\int^{\infty}_{\theta-w^{\prime}} n f(w^{\prime\prime})
dw^{\prime\prime} = n \left[ 1 - F(\theta - w^{\prime}) \right]
\label{eq:k_w_2}
\end{equation}
triangles. We obtain $C(k)$ by
the sum of \EQS(\ref{eq:k_w_1}) and (\ref{eq:k_w_2}) that
is weighted by the degree distribution. The normalization is given by
dividing it by $k(k-1)/2$ and by
another factor of 2, as each triangle is counted twice.
Consequently, we have for $w> \theta / 2$, or $k>
n\left[ 1- F\left(\theta / 2\right)\right]$
\begin{eqnarray}
C(k) &=& \frac{1}{2}
\frac{1}{k(k-1)/2}
\left\{ \int^{\infty}_w k f(w^{\prime})dw^{\prime} + 
\int^w_{\theta-w} n \left[ 1 - F(\theta - w^{\prime}) \right]
f(w^{\prime})dw^{\prime}
\right\}\nonumber\\
&=&\left\{ -1+2\frac{k}{n} + \left(1-\frac{k}{n}\right)
F\left( \theta-F^{-1}\left( 1-\frac{k}{n}\right) \right) \right.\nonumber\\
&&\left. -
\int^{k}_{n[1-F(\theta-F^{-1}(1-k/n))]}
\left(1-\frac{k^{\prime}}{n}\right) p(k^{\prime}) dk^{\prime} \right\}
\bigg/ \left(k/n\right)^2.
\label{eq:cluster1}
\end{eqnarray}
When $w\le \theta / 2$
or $k\le n\left[ 1- F\left(\theta / 2\right)\right]$,
we simply end up with
\begin{equation}
C(k) = 1.
\label{eq:cluster2}
\end{equation}
The vertices with $C(k) = 1$ forms the peripheral part of the network
that is connected to the cliquish core with smaller $C(k)$, as
schematically depicted in \FIG\ref{fig:separation}.  The core consists
of the vertices with $w\ge \theta / 2$, and it is similar to the
winner-take-all phenomenon found in growing network models
\cite{Bianconi,Krapivsky}. However, more than a single winners are
allowed in this model.  This core-peripheral separability is a
rigorous property of the threshold graph \cite{Golumbic}. It is also
consistent with the property of real networks that $C(k)$ saturates
for small $k$ \cite{Ravasz02,Ravasz03} and that vertices in the core
are densely connected \cite{ZhouMondragon}.

Regarding the network size, real networks are small with $L$
proportional to $\ln n$ or even less
\cite{SW,reviews,Barabasi99}. Our network has $L\le 2$ as far as it is
free from isolated vertices. This is because any pair of vertices can
be connected by a path of length 2 passing a vertex with a
sufficiently large weight is in the cliquish part (see
\FIG\ref{fig:separation}). One might ascribe this
ultrasmallness to the fact that the mean degree is of the order of
$n$, indicating too many edges per vertex.  However, the
mean degree can be kept finite by scaling $\theta$ according to the
increase in $n$, as discussed in \SEC\ref{sub:exponential}. It turns
out that this modification does not change $L$.

The correlation between the degrees of adjacent vertices also
characterizes networks \cite{reviews}.  Actually, degree
correlation can be positive or negative depending on the type of
network, as measured for real data using the degree of assortativity
\cite{Newman02prl}. Here we explain a simpler quantity to gain insight
into the degree correlation \cite{Klemm0202,Vazquez02,Vazquez03},
which was first analyzed in Ref. \cite{Boguna} for the present type of
network model.  We denote by $k_2$ the sum of the degrees of
$v$'s neighbors, which has degree $k$.  If the degree is
uncorrelated, $k_2/k$, or the average degree of the neighbors, is
independent of $v$ or $k$. For the threshold model, we derive
\begin{eqnarray}
k_2 &=& \int^{\infty}_{\theta-w} n f(w^{\prime})
n\left[ 1-F\left( \theta-w^{\prime} \right) \right] dw^{\prime}\nonumber\\
&=&
n^2 \left\{\frac{k}{n} - \int^n_{n[1-F(\theta-F^{-1}(1-k/n))]}
\left( 1-\frac{k^{\prime}}{n}\right) p(k^{\prime}) dk^{\prime}\right\}.
\label{eq:k_2}
\end{eqnarray}
Accordingly,
\begin{equation}
\frac{k_2}{k} = 
n\left\{ 1 - \frac{1}{k}\int^n_{n[1-F(\theta-F^{-1}(1-k/n))]}
(n-k^{\prime}) p(k^{\prime}) dk^{\prime}\right\},
\label{eq:corr2}
\end{equation}
which generally depends on $k$. 

\section{Examples}\label{sec:examples}

In this section, we 
calculate the quantities introduced in
\SEC\ref{sec:model} for some weight distributions $f(w)$.

\subsection{Exponential distribution}\label{sub:exponential}

\begin{figure}
\begin{center}
\includegraphics[height=2.25in,width=3.25in]{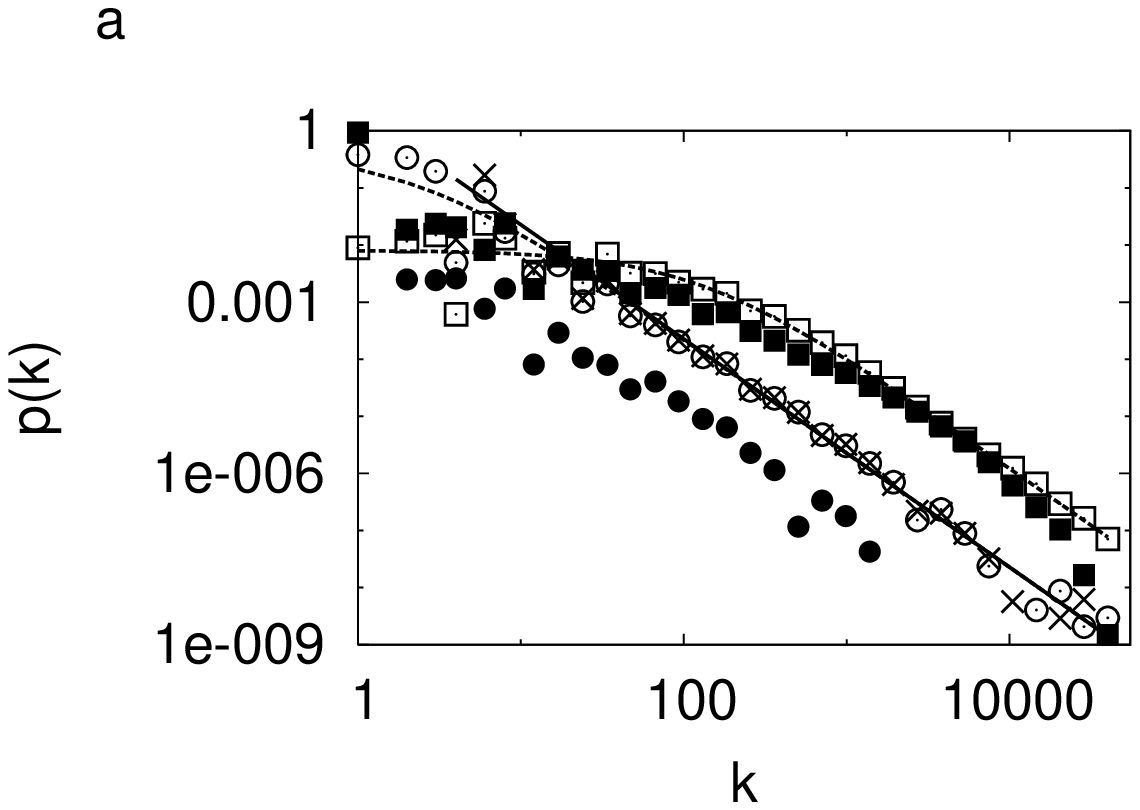}
\includegraphics[height=2.25in,width=3.25in]{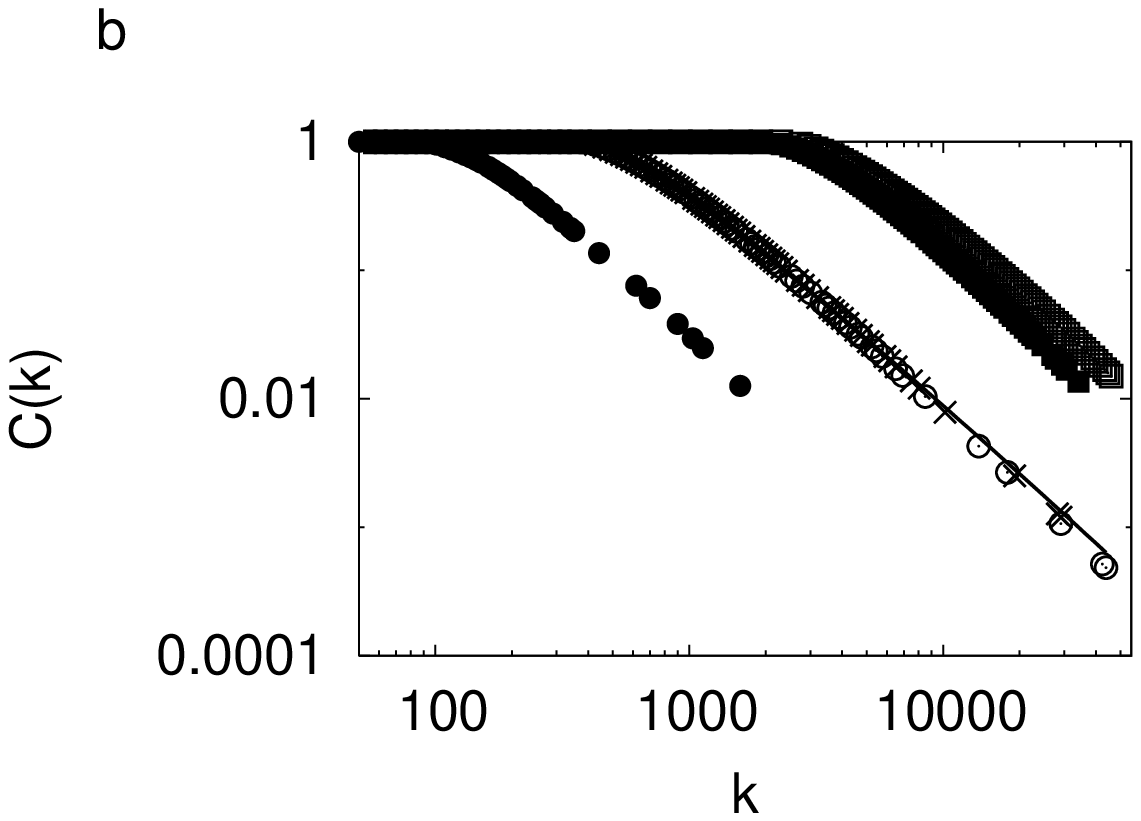}
\includegraphics[height=2.25in,width=3.25in]{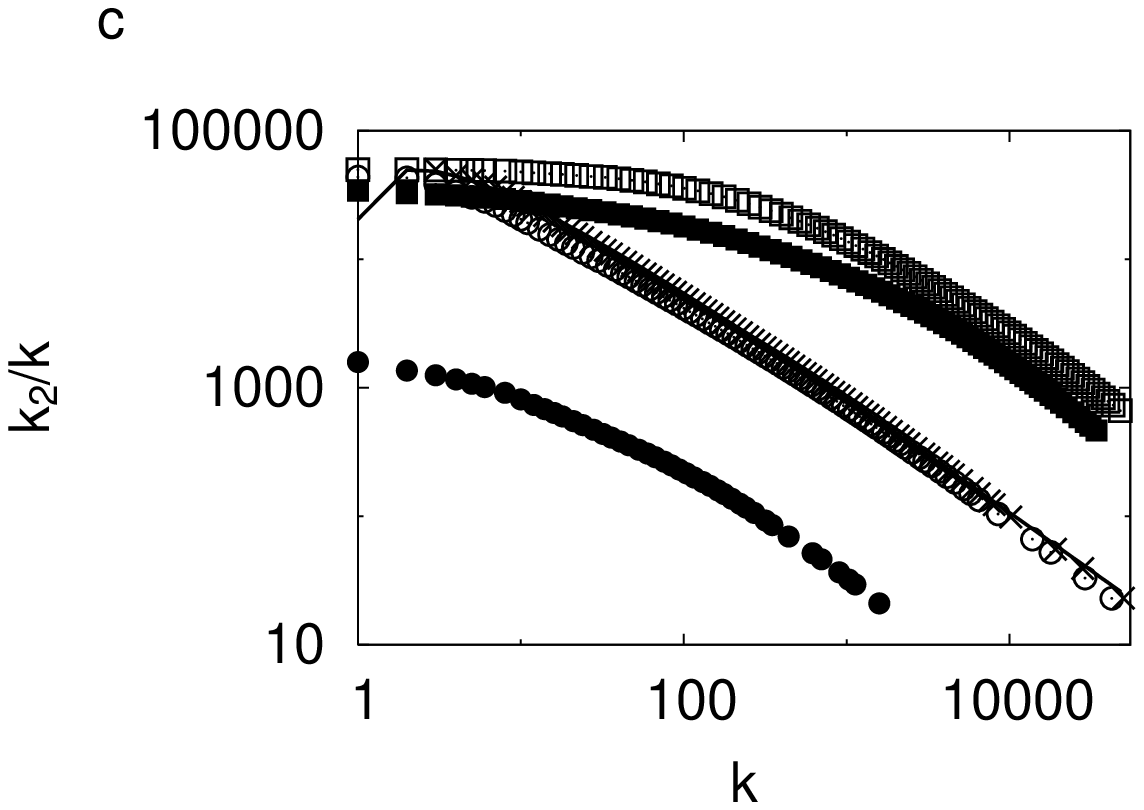}
\caption{}
\label{fig:explog}
\end{center}
\end{figure}

Let us begin with recapitulating 
the case of exponential weight distribution with deterministic
thresholding \cite{Caldarelli,Boguna}.
We set
\begin{equation}
f(w) = \lambda e^{-\lambda w} \quad
(0\le w).\label{eq:f(w)_exp}
\end{equation}
Assuming $\theta > 0$ so that
the generated networks are not trivial, for $w < \theta$,
combining \EQS(\ref{eq:p(k)}) and (\ref{eq:f(w)_exp})
results in
\begin{eqnarray}
p(k) &=& \frac{n e^{-\lambda \theta}}{k^2}\quad
(n e^{-\lambda\theta}\le k \le n).
\label{eq:p(k)_exp}
\end{eqnarray}
The scale-free distribution $p(k)\propto k^{-2}$ appears from random
weights whose distribution has nothing to do with power law, which is
a main claim of Ref. \cite{Caldarelli}.  A total of $\int^{\infty}_{\theta}
n f(w)dw = n {\rm e}^{-\lambda \theta}$ vertices are condensated at
$k=n$ and form a core \cite{Caldarelli,Boguna}.  In general, this sort
of condensation occurs when $f(w)$ has a lower cutoff.  However, the
number of vertices with $k=n$ can be made arbitrarily small by setting
large $\lambda\theta$, and this feature is not so essential.  We
numerically simulate a network with $n=50000$, which is fixed
throughout the paper, $\lambda=1$, and $\theta=10$.  Setting
$\lambda=1$ does not cause the loss of generality because only the
multiple of $\lambda$ and $\theta$ appears in \EQ(\ref{eq:p(k)_exp})
and the following quantities [\EQS(\ref{eq:C(k)_exp}) and
(\ref{eq:corr_exp})].  Figure~\ref{fig:explog}(a) shows that numerical
results (crosses) are predicted by \EQ(\ref{eq:p(k)_exp}) (lines)
sufficiently well \cite{Caldarelli,Boguna}.  In regard to clustering,
\EQ(\ref{eq:cluster1}) yields
\begin{equation}
C(k) = \left\{ \begin{array}{ll}
1 &
\quad (n e^{-\lambda\theta}\le 
k\le n e^{-\lambda\theta/2})\\
\frac{n^2}{k^2} e^{-\lambda\theta}
\left( 1+\lambda\theta+2\ln\frac{k}{n} \right) &
\quad (n e^{-\lambda\theta/2}< k \le n)
\end{array} \right.
\label{eq:C(k)_exp}
\end{equation}
which agrees with the numerical results in \FIG\ref{fig:explog}(b)
(crosses) (originally derived in Ref. \cite{Boguna}).
Equation~(\ref{eq:C(k)_exp}) shows that $C(k)$ nearly decays according
to the power law with exponent 2. However, analysis of real networks,
such as metabolic networks \cite{Ravasz02}, actor networks, semantic
networks, world wide webs, and the Internet \cite{Ravasz03}, suggests
$C(k)\propto k^{-1}$, which is also supported by some models
\cite{Vazquez02,Dorogovtsev02,Ravasz02,Ravasz03,Klemm0202,Vazquez03,Newman03,Szabo03}.
For this particular example, we have a larger scaling exponent.
Actually, the general power-law form $C(k)\propto
k^{-\gamma^{\prime}}$ ($\gamma^{\prime}>0$) is also reported in model
studies \cite{Vazquez02,Ravasz03,Szabo03} and in data analysis
\cite{Vazquez02}.  The clustering coefficient of the whole network is
\begin{equation}
C = \int^n_{n e^{-\lambda\theta}} C(k) p(k) dk
= 1 - \frac{4}{9} e^{-\lambda\theta/2} 
- \frac{5+3\lambda\theta}{9} e^{-2\lambda\theta}.
\end{equation}
In real networks, the average vertex degree denoted by $\left<
k\right>$ is independent of $n$ on a large scale
\cite{SW,reviews,Barabasi99}.  Since
\begin{equation}
\left< k \right> 
=  e^{-\lambda\theta} \left( n + \lambda\theta \right),
\end{equation}
finite $\left< k\right>$ is maintained by setting $\theta \cong
\lambda^{-1}\ln n$.  With this scaling of $\theta$, our model
produces a finite value of $C$ that does not decay to 0 in the limit
$n\to\infty$.  This result agrees with real data \cite{reviews}, and
$C$ is actually nonvanishing for more general $f(w)$.

Using \EQ(\ref{eq:corr2}), the average degree of neighbors
becomes, as shown in Ref. \cite{Boguna},
\begin{equation}
\frac{k_2}{k} = \frac{n^2{\rm e}^{-\lambda\theta}}{k}
\left( 1 + \lambda\theta + \ln \frac{k}{n}\right).
\label{eq:corr_exp}
\end{equation}
Since \EQ(\ref{eq:corr_exp}) is decreasing in $k$, the network is
disassortative \cite{Newman02prl} with negative degree
correlation, which is a property shared by some scale-free network models
\cite{Klemm0202,Vazquez02,Vazquez03} and some social and biological
networks \cite{Newman02prl,Vazquez02,reviews}.
More intuitively, disassortativity
of our model is a natural consequence of the core-peripheral structure.
Equation~(\ref{eq:corr_exp}) and \FIG\ref{fig:explog}(c) (crosses and
solid lines) show that there is an approximate scaling relation
$k_2/k\propto k^{-1}$ \cite{Vazquez03}.

\subsection{Logistic distribution}\label{sub:logistic}

The logistic distribution is often used, for example, in
statistics and economics.  Except the discrepancy in 
the asymptotic behavior, it
can be used as a more or less appropriate substitute for the Gaussian
distribution. The logistic distribution is more tractable
because it has an analytic form of
$F(w)$. For a given $\beta>0$, the logistic
distribution is defined by
\begin{equation}
f(w) = \frac{\beta e^{-\beta w}}
{\left( 1 + e^{-\beta w}\right)^2},\label{eq:f(w)_logistic}\\
\end{equation}
With
\begin{eqnarray}
F(w) &=& \frac{1}{1+ e^{-\beta w}},\quad (w\in {\bf R})\\
F^{-1}(x) &=& -\frac{1}{\beta} \ln \left( \frac{1}{x} - 1\right),
\label{eq:F-1(x)_logistic}\\
k &=& \frac{n e^{\beta(w-\theta)}}
{1 + e^{\beta(w-\theta)}},
\label{eq:k(w)_logistic}\\
w &=& \theta + \frac{1}{\beta}\ln \frac{k}{n-k}.
\label{eq:w(k)_logistic}
\end{eqnarray}
applied to \EQ(\ref{eq:p(k)}), we obtain
\begin{equation}
p(k) = \frac{n e^{-\beta\theta}}
{k^2 \left( 1 + e^{-\beta\theta}\frac{n-k}{k}\right)^2}
\quad (0\le k\le n).
\label{eq:p(k)_logistic}
\end{equation}
The power law $p(k)\propto k^{-2}$ is again manifested as $k$
approaches $n$.  If $1 \ll [(n-k)/k] e^{-\beta\theta}$, $k$
is relatively small, and $p(k)\cong n e^{\beta\theta}/(n-k)^2$
does not strongly depend on $k$. The
crossover from this regime to the power-law regime, which is found in
real data \cite{Ravasz02,Ravasz03} and derived by scaling ansatz
theory \cite{Szabo03}, occurs around $1 \cong [(n-k)/k]
e^{-\beta\theta}$, or $k\cong [n/(e^{\beta\theta}+1)]$.  A
larger value of $\beta\theta$ provides a wider range of $k$ in which
the power law holds.  In this range, \EQ(\ref{eq:f(w)_logistic}) is of
course approximated by the exponential distribution represented by
\EQ(\ref{eq:f(w)_exp}) with $\beta = \lambda$.  For $k$ small relative
to $n /(e^{\beta\theta}+1)$, $f(w)$, with a correspondingly small $w$,
does not decay exponentially or
even monotonically.  Therefore, when $k$ is small, the number of
vertices with degree $k$ is not large enough to support the power law.

We compare in \FIGS\ref{fig:explog}(a) the numerical results for
$\theta = 6$ (open squares) and $\theta = 10$ (open circles) with the
corresponding theoretical results in \EQ(\ref{eq:p(k)_logistic})
(dotted lines).  We have set $\beta = 1$ without losing
generality for the same reason as in \SEC\ref{sub:exponential}. The
effect of $\theta$ on the position of crossover is clear in the
figure. Since the integrals in \EQS(\ref{eq:cluster1}) and
(\ref{eq:corr2}) cannot be explicitly calculated, numerically
evaluated $C(k)$ and $k_2/k$ with the same parameter values are shown
in \FIGS\ref{fig:explog}(b) and \ref{fig:explog}(c),
respectively. Similar to the case
of the exponential distribution, the crossover from the plateau to the
power law is observed for $C(k)$ with the same scaling exponent
$C(k)\propto k^{-2}$. Also with regard to the degree correlation,
$k_2/k\propto k^{-1}$ approximately holds except for small $k$.

\subsection{Gaussian distribution}\label{sub:gaussian}

The Gaussian distribution can be a standard null hypothesis on the
weight distribution. Since it does not have the analytical form of
$F^{-1}(x)$, we perform straightforward numerical simulations with
$\theta = 6$ and $\theta = 10$ to examine $p(k)$, $C(k)$, and $k_2/k$.
The Gaussian distribution is assumed to have mean 0 and standard
deviation $1.7$ to roughly approximate the logistic distribution with
$\beta = 1$, which has been used in \SEC\ref{sub:logistic}.  In spite
of different asymptotic decay rates of $f(w)$, \FIG\ref{fig:explog}
indicates that $p(k)$, $C(k)$, and $k_2/k$ for the Gaussian
distribution do not differ so much from those for the logistic
distribution, disregarding the crossover points. This implies a rather
universal existence of power law behavior, which is discussed in more
detail in \SEC\ref{sec:universality}.

\subsection{Pareto distribution}\label{sub:pareto}

\begin{figure}
\begin{center}
\includegraphics[height=2.25in,width=3.25in]{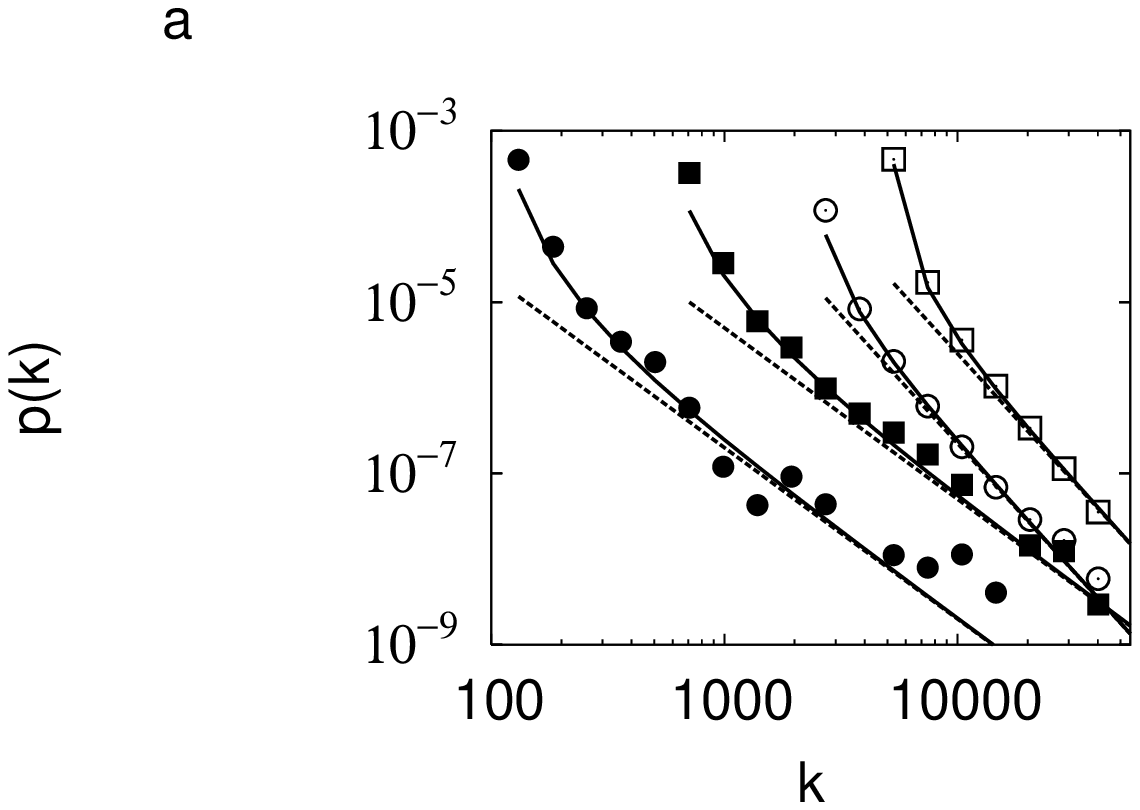}
\includegraphics[height=2.25in,width=3.25in]{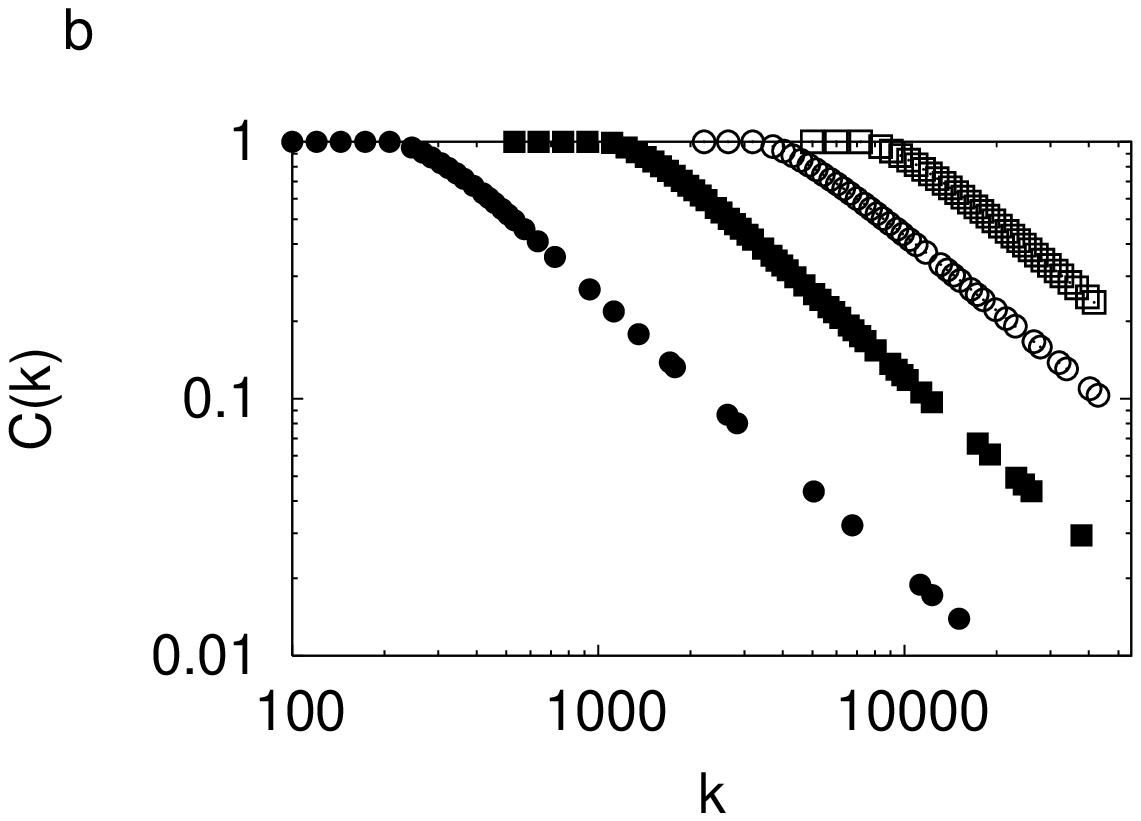}
\includegraphics[height=2.25in,width=3.25in]{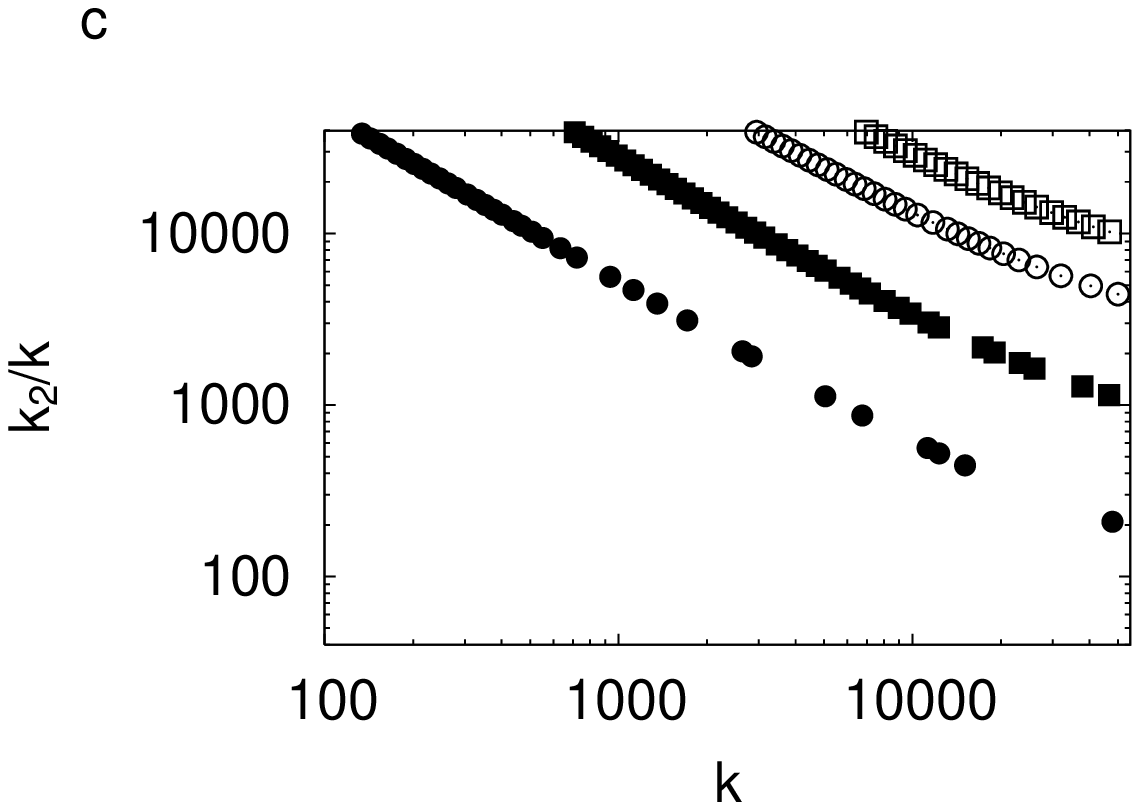}
\caption{}
\label{fig:par}
\end{center}
\end{figure}

The Pareto distribution, which is equipped with an inherent power law,
is often observed in, for example, distributions of capitals and
company sizes \cite{Pareto}.  It is defined by
\begin{equation}
f(w) = \frac{a}{w_0}\left( \frac{w_0}{w}\right)^{a+1}
\quad (w\ge w_0),
\end{equation}
where $a>0$ and $w_0>0$. Nontrivial networks
form if we choose $\theta>2 w_0$.
We obtain
\begin{eqnarray}
F(w) &=& 1 - \left( \frac{w_0}{w} \right)^a\quad (w\ge w_0),
\label{eq:F(w)_pareto}\\
F^{-1}(x) &=& \frac{w_0}{(1-x)^{1/a}}.
\label{eq:F-1(w)_pareto}
\end{eqnarray}
When $w\le \theta-w_0$, it is straightforward to derive
\begin{eqnarray}
k &=& n \left( \frac{w_0}{\theta - w}\right)^a,\quad
\left( n\left( \frac{w_0}{\theta-w_0}\right)^a \le k < n\right)\\
w &=& \theta - \left( \frac{n}{k} \right)^{1/a} w_0,\\
p(k) &=& \frac{n^{1/a}}
{\left( \frac{\theta}{w_0}k^{1/a} - n^{1/a} \right)^{a+1}}.
\label{eq:p(k)_pareto}
\end{eqnarray}
A total of
\begin{equation}
\int^{\infty}_{\theta-w_0}nf(w)dw =n \left(
\frac{w_0}{\theta-w_0}\right)^a
\end{equation}
vertices with $w\ge \theta-w_0$ are condensated at
$k=n$.  When $n(w_0/\theta)^a\ll k < n$, $p(k)$ can be approximated by
a power law 
\begin{equation}
p(k) \cong \left(\frac{w_0}{\theta}\right)^{a+1}n^{1/a}
k^{-(a+1)/a}.
\label{eq:p(k)_pareto_approx}
\end{equation}
By modulating $a$, we can produce a scale-free $p(k)$ with arbitrary
$\gamma=(a+1)/a>1$. An observed $\gamma$ in turn serves to estimate
$a$ and $f(w)$, which may underly, for example, fractal dynamics of
economical quantities, as well as network formation.  The scaling
exponent for $p(k)$ differs from that for $f(w)$, and a faster decay
of $f(w)$ with a larger $a$ yields a slower decay of
$p(k)$.  Numerical results for $p(k)$ are shown in
\FIG\ref{fig:par}(a) with $w_0=1$.  We set $(a,\theta)=(0.5, 100)$
(open squares) and $(0.5, 500)$ (open circles), yielding $\gamma = 3$,
and $(a,\theta)=(1, 100)$ (closed squares) and $(1, 500)$ (closed
circles), resulting in $\gamma = 2$. The results are consistent with
the theoretical prediction based on \EQ(\ref{eq:p(k)_pareto}) (solid
lines) and also with \EQ(\ref{eq:p(k)_pareto_approx})
(dotted lines).

Substituting \EQS(\ref{eq:F(w)_pareto}), (\ref{eq:F-1(w)_pareto}),
and (\ref{eq:p(k)_pareto}) into \EQS(\ref{eq:cluster1}),
(\ref{eq:cluster2}), and (\ref{eq:corr2}), respectively, yields
\begin{eqnarray}
C(k) &=&
\frac{n}{k} 
+ \left(\frac{n}{k}-\frac{n^2}{k^2}\right) \left(\frac{w_0}{\theta - w_0
  \left(\frac{n}{k}\right)^{1/a}}\right)^a 
- 
\frac{n^2 a}{k^2} \left( \frac{w_0}{\theta} \right)^a
\int^{\left(k/n\right)^{1/a}(\theta/w_0)}_{\theta/[\theta
      - w_0(n/k)^{1/a}]} 
\frac{x^{a-1} - \left(\frac{w_0}{\theta}\right)^a
  x^{2a-1}}{(x-1)^{a+1}} dx
\nonumber\\
&=& 
\frac{n}{k} \left(\frac{w_0}{\theta - w_0
  \left(\frac{n}{k}\right)^{1/a}}\right)^a 
+
\frac{n^2 a}{k^2} \left( \frac{w_0}{\theta} \right)^{2a}
\int^{1-\left(n/k\right)^{1/a}(w_0/\theta)}_{\left(n/k\right)^{1/a}
(w_0/\theta)} y^{-a}(1-y)^{-a-1}dy,\nonumber\\
&& \quad \left(k > n\left( \frac{2 w_0}{\theta}\right)^a\right),
\label{eq:C(k)_pareto1}
\end{eqnarray}
\begin{equation}
C(k) = 1,\quad \left(n\left( \frac{w_0}{\theta-w_0}\right)^a\le
k\le n\left( \frac{2 w_0}{\theta}\right)^a\right)
\end{equation}
and
\begin{eqnarray}
\frac{k_2}{k} &=& 
n
\left\{ 1 - \frac{na}{k}
\left(\frac{w_0}{\theta} \right)^a
\int^{\theta/w_0}_{\theta/[\theta-w_0(n/k)^{1/a}]}
\frac{x^{a-1} - \left(\frac{w_0}{\theta}\right)^a x^{2a-1}}{(x-1)^{a+1}} dx
\right\}
\nonumber\\
&=& \frac{n^2}{k}\left( \frac{w_0}{\theta-w_0}\right)^a
+ \frac{n^2 a}{k}\left( \frac{w_0}{\theta}\right)^{2a}
\int^{1-\left(n/k\right)^{1/a}(w_0/\theta)}_{w_0/\theta} y^{-a}(1-y)^{-a-1}dy,
\label{eq:deg_corr_pareto}
\end{eqnarray}
where we set $y=x^{-1}$.
The integral in \EQ(\ref{eq:C(k)_pareto1}) is nonnegative and
does not depend so much on $k$ when $k$ tends large.
Therefore, 
$C(k)\propto k^{-1}$ is expected based on the first term, which  
is consistent with the numerical
results for $k$ larger than the crossover value $k\cong n\left(
2 w_0/\theta\right)^a$ [\FIG\ref{fig:par}(b)].
The scaling law $C(k)\propto k^{-1}$,
as opposed to $C(k)\propto k^{-2}$
for the exponential $f(w)$, rather agrees
with real data \cite{Ravasz02,Ravasz03}.

Similarly, \EQ(\ref{eq:deg_corr_pareto}) and 
the simulation results shown in
\FIG\ref{fig:par}(c) suggest $k_2/k\propto k^{-1}$
for a sufficiently large $k$.
Equations~(\ref{eq:C(k)_pareto1}) and
(\ref{eq:deg_corr_pareto}) show that the scaling exponents 
of both $C(k)$ and $k_2/k$ do not depend on $\gamma$ or $a$.

\subsection{Cauchy distribution}\label{sub:cauchy}

For the Cauchy distribution
\begin{equation}
f(w) = \frac{1}{\pi (1+w^2)}\quad (w\in {\bf R}),
\end{equation}
we obtain
\begin{eqnarray}
F^{-1}(x) &=& \tan \frac{\pi}{2} \left(2x-1\right),\\
w &=& \theta - \tan \frac{\pi}{2}\left(1-\frac{2k}{n}\right),\\
p(k) &=&
 \frac{1}{n} \frac{1+\tan^2 \frac{\pi}{2}\left( 1-\frac{2k}{n}\right)}
{1+\left(\theta -\tan
\frac{\pi}{2}\left(1-\frac{2k}{n}\right)\right)^2}
\quad (0\le k\le n).
\label{eq:p(k)_cauchy}
\end{eqnarray}

\begin{figure}
\begin{center}
\includegraphics[height=2.25in,width=3.25in]{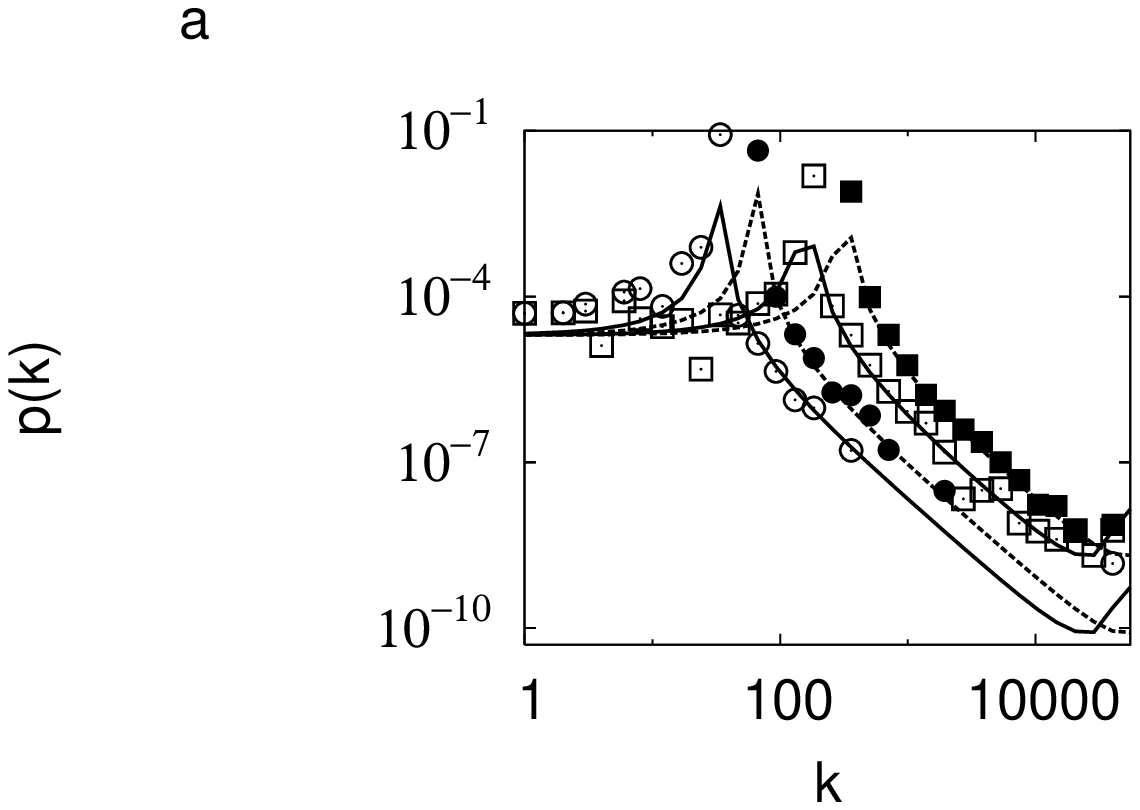}
\includegraphics[height=2.25in,width=3.25in]{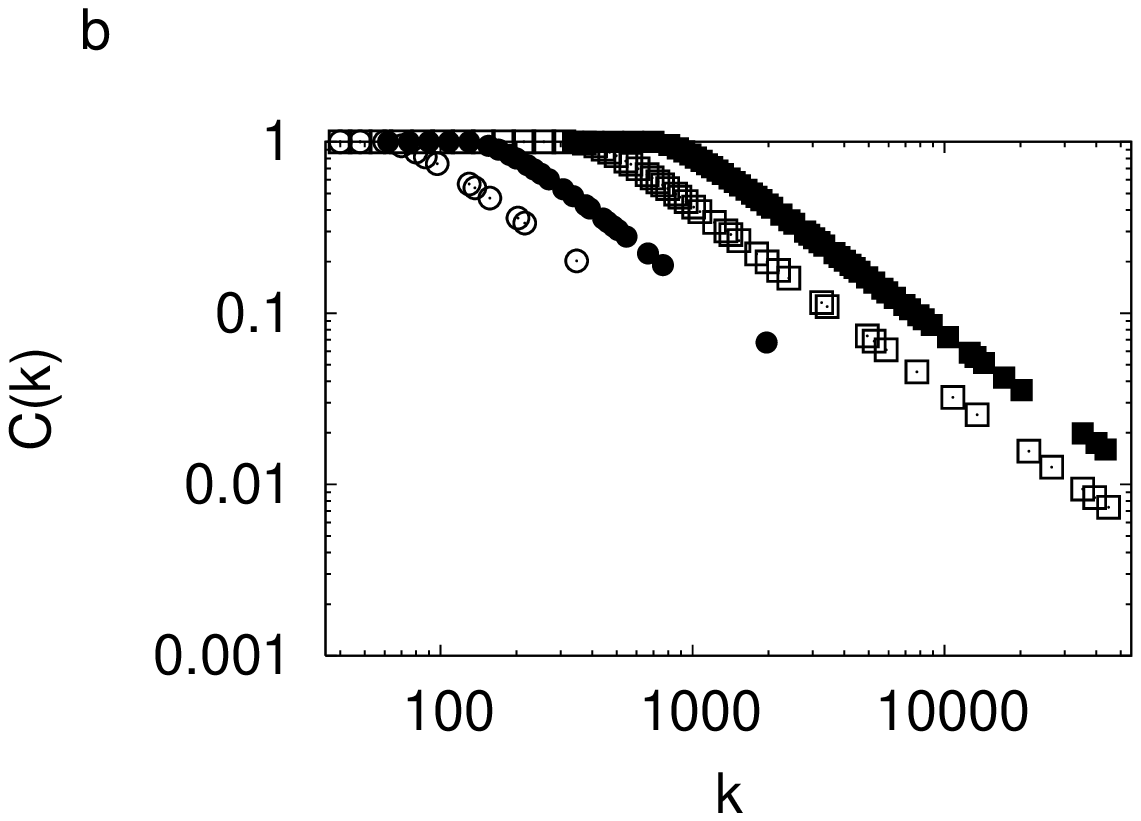}
\includegraphics[height=2.25in,width=3.25in]{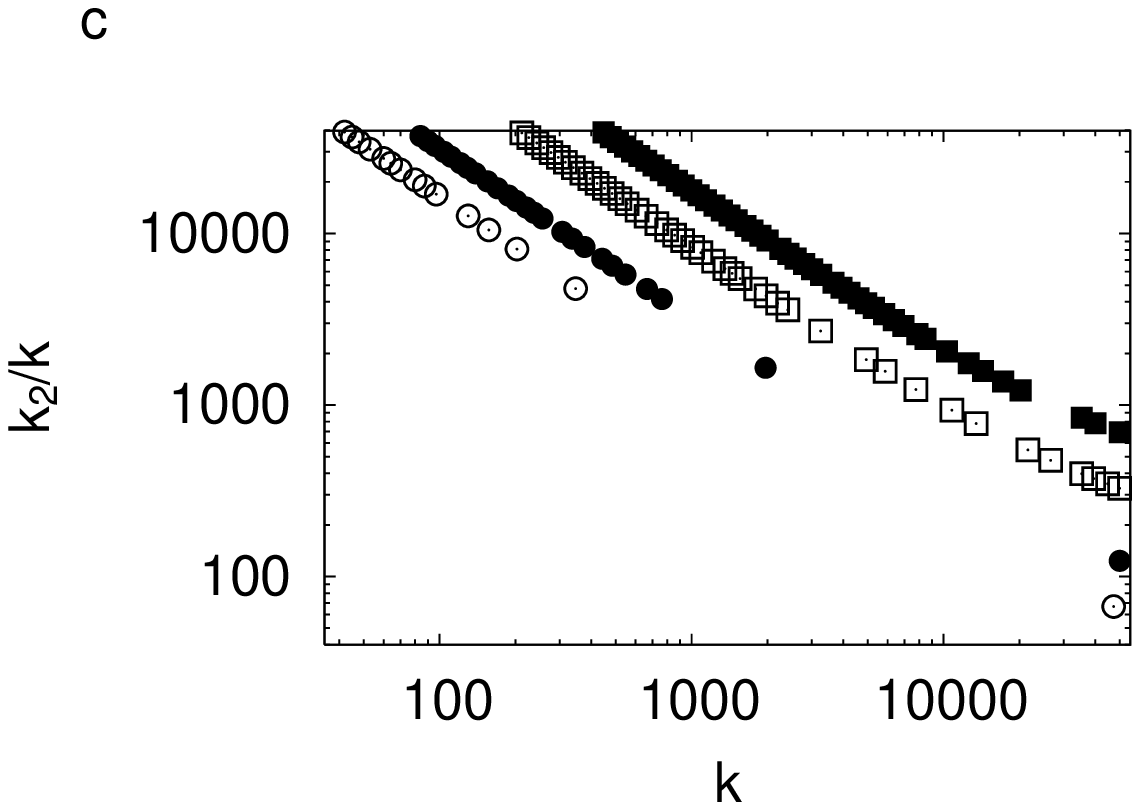}
\caption{}
\label{fig:cau}
\end{center}
\end{figure}

Numerically obtained $p(k)$, $C(k)$, and $k_2/k$ together
with \EQ(\ref{eq:p(k)_cauchy}) are shown in \FIG\ref{fig:cau} for
$\theta = 100$ (open squares) and for $\theta = 500$ (open circles).
According to \EQ(\ref{eq:p(k)_cauchy}), the monotonicity of $p(k)$ is
marred because $p(0)= p(n) = 1/n$.  A particular choice of
$\theta=0$ even gives rise to the uniform $p(k)$. For general
$\theta$, however, $p(k)$ has the unique maximum and minimum between
$k=0$ and $k=n$ as shown in \FIG\ref{fig:cau}(a) by solid lines.
Existence of the characteristic vertex degree corresponding to the
peak of $p(k)$ is a feature shared by random, regular and
small-world networks \cite{SW,reviews}. The peak appears because of
the unimodality of $f(w)$, which yields the plateaus of $p(k)$ in the
case of the logistic and Gaussian distributions (see
\SECS\ref{sub:logistic} and \ref{sub:gaussian}).  Nonetheless, as for
the Pareto distribution with $a=1$, which has the same asymptotics
$f(w)\propto w^{-2}$ as the Cauchy distribution, approximate power
laws with $\gamma=2$ are observed for intermediate values of $k$.

The one-sided Cauchy distribution on the half line facilitates a fairer
comparison with the Pareto distribution that also has the lower cutoff
of $w$. We define the one-sided Cauchy distribution by
\begin{equation}
f(w) = \frac{2}{\pi (1+w^2)} \quad (w\ge 0).
\end{equation}
Then it holds that $F^{-1}(x) = \tan (\pi/2)x$ and
\begin{equation}
p(k) = \frac{2}{n} \frac{1+\tan^2 \frac{\pi}{2}\left( 1-\frac{k}{n}\right)}
{1+\left(\theta -\tan
\frac{\pi}{2}\left(1-\frac{k}{n}\right)\right)^2}
\quad (0\le k\le n).
\label{eq:p(k)_one}
\end{equation}
Figure~\ref{fig:cau}(a) shows the numerical results for $p(k)$ with
$\theta=100$ (closed squares) and $\theta=500$ (closed circles),
accompanied by the prediction by \EQ(\ref{eq:p(k)_one}) (dotted
lines). The approximate power law holds even for $k$ close to $n$,
which contrasts with the case of the standard Cauchy distribution.
The behavior of $C(k)$ and $k_2/k$ shown in \FIGS\ref{fig:cau}(b) and
\ref{fig:cau}(c), respectively, resembles that for the Pareto and standard Cauchy
distributions.  In \SECS\ref{sub:exponential} and \ref{sub:logistic},
we have inspected consequences of using exponential
$f(w)$ and logistic $f(w)$. Including the comparison between
the Pareto and Cauchy distributions examined here, effect of a lower
cutoff of $f(w)$ does not seem so prominent.

\section{Why power law with $\gamma=2$?}\label{sec:universality}

\begin{figure}
\begin{center}
\includegraphics[height=2.25in,width=3.25in]{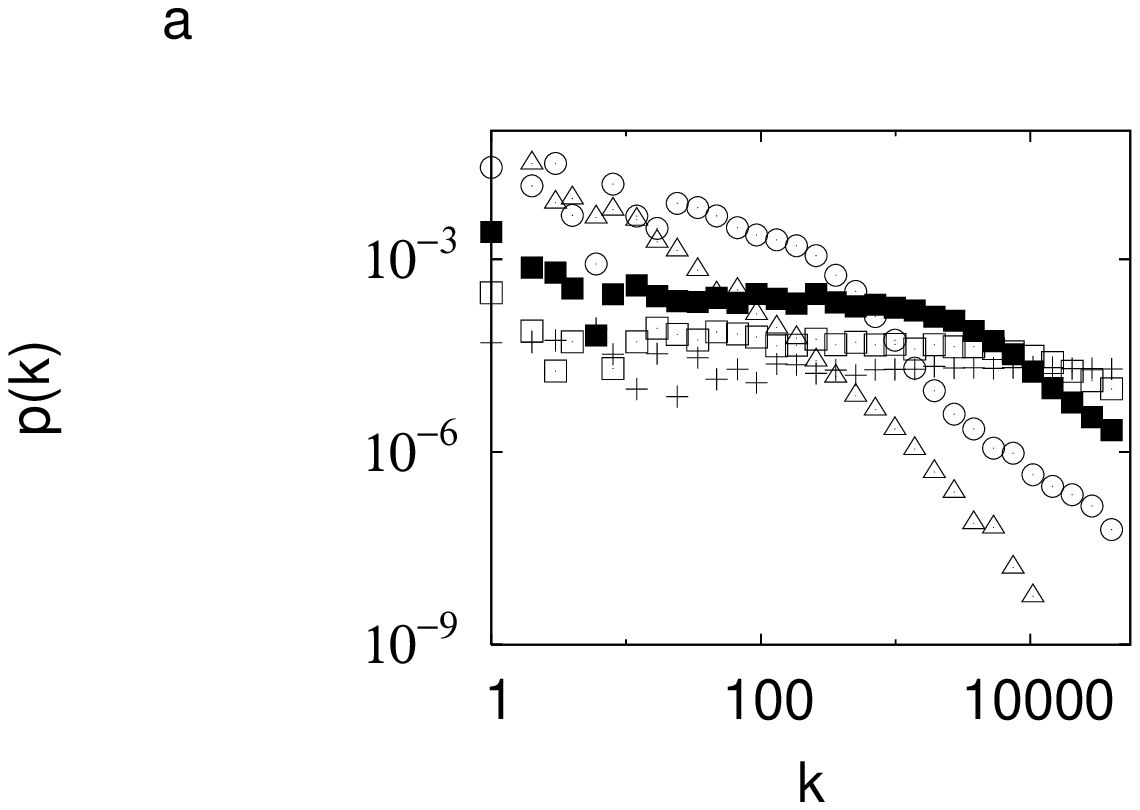}
\includegraphics[height=2.25in,width=3.25in]{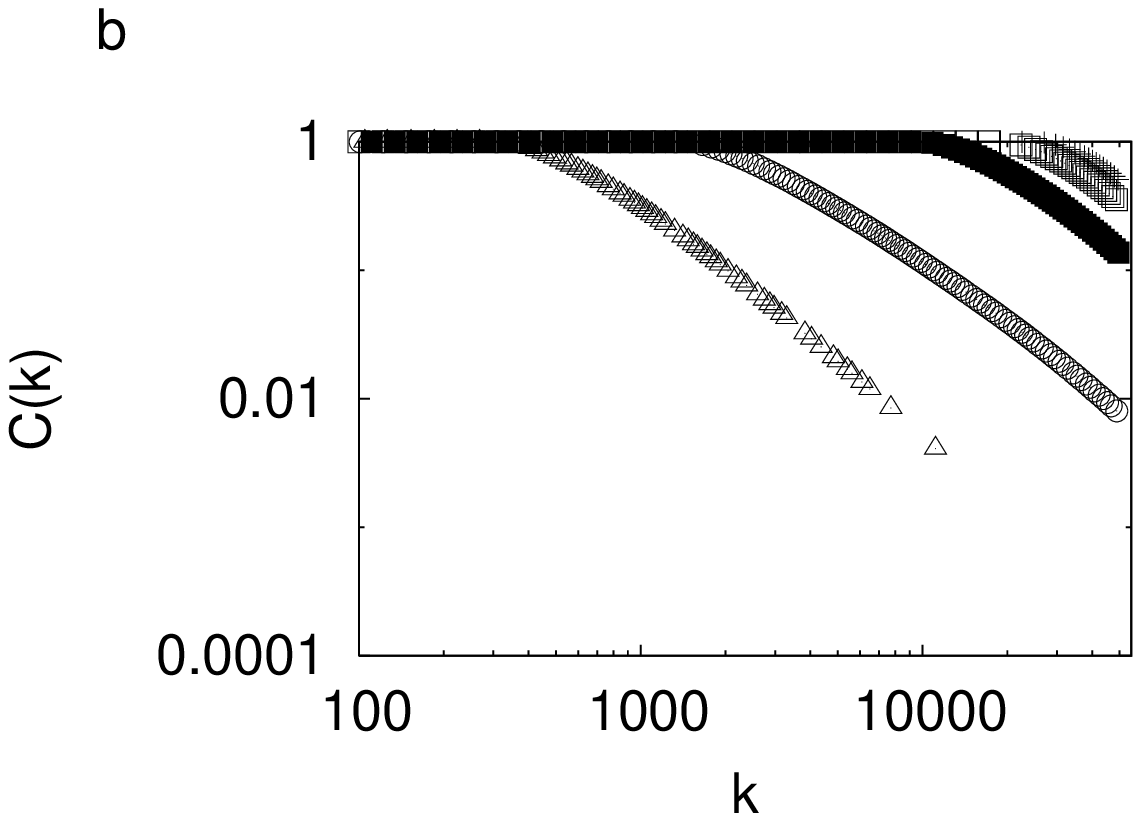}
\includegraphics[height=2.25in,width=3.25in]{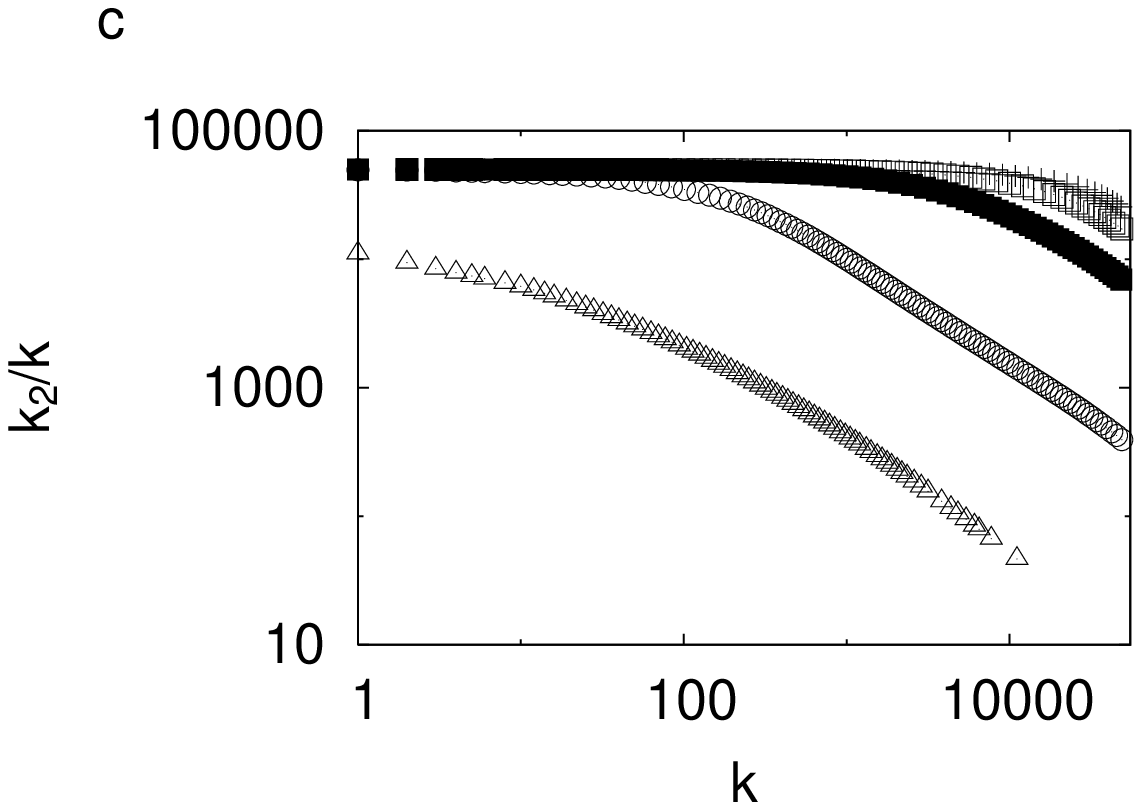}
\caption{}
\label{fig:evolve}
\end{center}
\end{figure}

The power law of $p(k)$ with $\gamma=2$ seems universal for
thresholding mechanisms not only because a wide class of $f(w)$
generates it but also owing to its stability. 
To be more specific,
weights can be
the vertex degrees themselves, as is implied by the BA model. Indeed,
$k$ can represent how central or influential a node is \cite{Barrat}.
Then let us iterate our construction algorithm to simulate an evolving
but not growing network with dynamic $f(w)$ and $p(k)$.
Let us simulate a dynamical network with $n=50000$.
Initially, $w$ is uniformly distributed on $[0,1]$, and the
thresholding algorithm determines $k$. Then we set $w=k/n+\xi$, where
$\xi$ is the Gaussian white noise with standard deviation
$\sigma=0.2$, and iterate the dynamics. The numerical results are
shown in \FIG\ref{fig:evolve} with $\theta = 1$.  In the early stages
[crosses in \FIG\ref{fig:evolve}(a)], $k$ is distributed more or less
uniformly since a uniform $f(w)$ yields a uniform $p(k)$ just
accompanied by a singularity at $k=0$ or $k=n$, which is easily
checked with \EQ(\ref{eq:p(k)}). As the iteration goes on, however,
$p(k)$ converges to a power law with $\gamma=2$. Similarly,
$C(k)\propto k^{-1}$ and $k_2/k\propto k^{-1}$ are eventually realized
as shown in \FIGS\ref{fig:evolve}(b) and \ref{fig:evolve}(c).
Although $\theta$ too far from 1 or excessively small $\sigma$ results
in a complete or totally disconnected network, $p(k)\propto k^{-2}$
emerges robustly against changes in $\theta\ge 1$ and $\sigma>0.15$
unless noise is not extremely large. 

By the analogy of the Pareto case, if a power law with
$\gamma\neq 2$ is obtained, $\gamma$ will be transformed by the map
$\gamma=a+1 \to \gamma=(a+1)/a$, namely, $\gamma \to
\gamma/(\gamma-1)$. This map has a unique positive fixed
point $\gamma=2$. Actually, the map is neutrally stable at $\gamma=2$ with
eigenvalue $-1$, which implies oscillation.
This argument does not directly support but may underlie
the emergence of $p(k)\propto k^{-2}$.
In addition, $p(k)$ converges to
$p(k)\propto k^{-2}$ stably with respect to the choice of initial
distribution $f(w)$. It is in a striking contrast with the case of
competitive growing networks with vertex weights, which generate
$p(k)\propto k^{-\gamma}$ only for a limited class of weight
distributions \cite{Bianconi}.  More broadly, general cooperative
networks in which interactions of multiple vertices leads to
interconnection \cite{Caldarelli} may have stable power laws, possibly
with $\gamma\neq 2$. As an
example, we can assume $w$ of the next generation to be proportional 
to $k^{x}$ ($x> 0$),
which is often the case in real networks \cite{Barrat}. In
this case, it is easy to show that the
neutrally stable fixed point of the abovementioned map
is $\gamma=1+\sqrt{x}$.  Our model,
which yields $\gamma=2$, sets a baseline example of this class.

In terms of clustering, we have observed the discrepancy in the
scaling law $C(k)\propto k^{-2}$ for the exponential, logistic, and Gaussian
distributions with $C(k)\propto k^{-1}$ for the Pareto, Cauchy, and
one-sided Cauchy distributions.
The Pareto distribution
with $a=1$ results in $p(k)\propto k^{-2}$,
which coincides with the scaling law 
for the exponential type of $f(w)$.
It means that the consistency in $p(k)$ for
different choices of $f(w)$ does not necessarily mean the
consistency in network structure. The difference
between the exponential tail and the power-law tail of $f(w)$ is likely 
to cause qualitative discrepancy in $C(k)$.
On the other hand, the degree correlation behaves
similarly in all the examined cases, namely, $k_2/k\propto k^{-1}$.
From a dynamical point of view,
$C(k)$ evolves from a general form to
$C(k)\propto k^{-1}$, which is realistic
\cite{Ravasz02,Ravasz03}.
This scenario matches the simulated dynamics
shown in \FIG\ref{fig:evolve}(b). In this case, $C(k)$
is initially just large for most vertices and
converges to $C(k)\propto k^{-1}$, which
reflects the eventual separation of the network into
the core and the peripheral part.

Boosted by the original BA models, the power law of $p(k)$ with
$\gamma = 3$ has been pronounced in the first place
\cite{reviews,Barabasi99}. Moreover, in percolation and contact
processes on scale-free networks, the
critical value of the infection rate is extinguished if $\gamma\le 3$
\cite{epidemic}. These results suggest relevance of the power law with
$\gamma\cong 3$.  However, real scale-free networks have more
dispersed values of $\gamma$ \cite{reviews}, and many models have been
proposed so that $\gamma$ is tunable somewhere between 2 and $\infty$
\cite{Albert00prl,Krapivsky,Bianconi,Barabasi01,Dorogovtsev02,Ravasz02,Ravasz03,Klemm0202,Barrat,Holme,Jung,Volchenkov}.
In contrast, the present model with intrinsic vertex weights (also see
Refs. \cite{Caldarelli,Boguna}) and another type of thresholding model
\cite{Volchenkov} broadly yield $\gamma=2$. We speculate that
$\gamma=2$ is another general law. In the parameter space of $\gamma$,
$\gamma=2$ as well as $\gamma=3$ often emerges as phase
transition points of network characteristics
\cite{Albert00prl,Krapivsky,Dorogovtsev02,Chung,Barrat}. The
$\gamma=2$ law may be common to cooperative models such as those with
thresholding, while the $\gamma=3$ law underlies the
competitive models represented by network growth with preferential
attachment.  Actually, many real networks in the scale-free regime
have $\gamma$ close to 2 rather than to 3 \cite{reviews}. Some
networks such as the world wide web, e-mail networks, language
networks, and ecological networks have $\gamma$ even smaller than 2
\cite{smaller_than_2}. Some of these observations can be understood as
small deviations from our $\gamma = 2$ law, which may be explained by
proper modification of the model \cite{Caldarelli,Boguna}, for
example, by introducing stochastic thresholding
\cite{Chung,Newman_Goh}, nonlinear relations between $k$ and $w$
\cite{Barrat}, or many-body interactions.

\section{Conclusions}

We have shown that the thresholding model, which is in the class of networks
with intrinsic
vertex weights, generates scale-free
networks with $\gamma=2$, large $C$, and small $L$ for a broad choice
of weight distributions.  Even if we start with an arbitrary weight
distribution, $p(k)\propto
k^{-2}$ and $C(k)\propto k^{-1}$ are finally obtained.
The competitive mechanisms, such as network growth with
preferential attachment or hierarchical structure, are not mandatory
for generating realistic networks \cite{Caldarelli}.
The cooperative thresholding mechanisms also result in desired
properties rather generally, and they yield somewhat different
characteristics from those of growing types of networks.
In addition, they allow plausible physical
interpretations, have core-peripheral structure,
are equipped with inhomogeneity as in real networks,
and facilitate analytical calculations \cite{Caldarelli,Boguna}.

\begin{acknowledgements}
We thank G. Caldarelli, P. De Los Rios, and A. Flammini for valuable
comments and 
introducing us the important references.
This study is supported by the Grant-in-Aid for Scientific Research,
Grant-in-Aid for Young Scientists (B) (Grant No. 15700020),
and the Grant-in-Aid for Scientific Research (B)
(Grant No. 12440024) of Japan Society of the Promotion of Science.
\end{acknowledgements}

\newpage

Figure captions

\bigskip

Figure 1: Schematic diagram of the threshold model.

\bigskip

Figure 2: Numerical results for (a) $p(k)$, (b) $C(k)$, and (c)
$k_2/k$ using the networks of size $n=50000$ generated by
thresholding.  The weight functions are taken to be exponential with
$\lambda = 1$, $\theta = 10$ (crosses), logistic with $\beta=1$,
$\theta = 6$ (open squares) and $\beta=1$, $\theta=10$ (open circles),
Gaussian (mean 0 and standard deviation $1.7$) with $\theta=6$ (closed
squares), and Gaussian with $\theta=10$ (closed circles).  The
theoretical predictions are shown for the exponential distribution
(solid lines) and the logistic distributions (dotted lines).

\bigskip

Figure 3: Numerical results for (a) $p(k)$, (b) $C(k)$, and (c)
$k_2/k$ for the Pareto weight distributions with $n=50000$ and
$w_0=1$. We set $a=0.5$, $\theta = 100$ (open squares), $a=0.5$,
$\theta = 500$ (open circles), $a=1$, $\theta = 100$ (closed squares),
and $a=1$, $\theta = 500$ (closed circles).  In (a), $p(k)$ estimated
by \EQ(\ref{eq:p(k)_pareto}) and the by power law approximation in
\EQ(\ref{eq:p(k)_pareto_approx}) are also shown with solid lines and
dotted lines, respectively.

\bigskip

Figure 4: Numerical results for (a) $p(k)$, (b) $C(k)$, and (c)
$k_2/k$ for the Cauchy weight distribution with $n=50000$ with
$\theta = 100$ (open
squares), $\theta = 500$ (open circles), and the one-sided Cauchy
distribution with $\theta = 100$ (closed squares), and $\theta = 500$
(closed circles). In (a), the analytically estimated $p(k)$
for the Cauchy distribution [\EQ(\ref{eq:p(k)_cauchy})] and 
for the one-sided Cauchy distribution [\EQ(\ref{eq:p(k)_one})]
are also shown by solid and dotted lines, respectively.

\bigskip

Figure 5: The evolution of (a) $p(k)$, (b) $C(k)$, and (c) $k_2/k$
with the repetitive thresholding. We set
$n=50000$ and $\theta = 1$.
The data shown are those
after 1 (crosses), 8 (open squares), 10 (closed squares),
12 (circles), and 15 (triangles) rounds.

\clearpage

\clearpage

\clearpage

\clearpage

\clearpage

\end{document}